\documentclass[5p]{elsarticle}

\usepackage{graphicx, amssymb,verbatim,amsmath,eepic,mathtools}

\newtheorem{theorem}{Theorem}[section]
\newtheorem{remark}{Remark}

\newcommand{\Sab}{ S_{\alpha, \beta} }
\newcommand{\sab}{ s_{\alpha, \beta} }

\newcommand{\Cab}{ C_{\alpha, \beta} }
\newcommand{\Cba}{ C_{\beta, \alpha} }

\newcommand{\ablim}{\lim\limits_{\substack{\alpha \to 1 \\ \beta \to 1}}}

\newcommand{\set}{\mathbb}

\newcommand{\ra}{\rightarrow}

\newcommand{\sR}{\mathbb R}
\newcommand{\sN}{\mathbb N}

\begin{document}
%%%%%%%%%%%%%%%%%%%%%%%%%%%%%%%%%%%%%%%%%%

\begin{frontmatter}
\journal{Physics Letters A}

%\title{On the generalization of Shannon-Khinchin Axioms and \\ the Uniqueness Theorem for the Nonextensive Entropy}

\title{Comments on ``A two-parameter generalization of \\ Shannon-Khinchin Axioms and the uniqueness theorem''}
%%%
\tnotetext[t1]{This research was supported by the Ministry of Education, Republic of Serbia, Grant No. III44006}

\author[misanu]{Velimir M. Ili\'c}
\ead{velimir.ilic@gmail.com}

\author[misanu]{Edin H. Mulali\'c\corref{cor}}
\ead{edinmulalic@yahoo.com}

\author[fosun]{Miomir S. Stankovi\'c}
\ead{miomir.stankovic@gmail.com}

\cortext[cor]{Corresponding author. Tel.: +381649908213}
\address[misanu]{Mathematical Institute of the Serbian Academy of Sciences and Arts, Kneza Mihaila 36, 11000 Beograd, Serbia}
\address[fosun]{University of Ni\v s, Faculty of Occupational Safety, \v Carnojevi\'ca 10a, 18000 Ni\v s, Serbia}
%%%%%%%%%%%%%%%%%%%%%%%%%%%%%%%%%%%%%%%%%%%%%%%%%%

\begin{abstract}
Wada and Suyari proposed a two-parameter
generalization of Shannon-Khinchin axioms (TGSK axioms) [\textit{T. Wada and H. Suyari, Physics
Letters A, 368(3)}]. We derive a new class of entropies which differs from Wada-Suyari's class by fixing the incorrectness which occurs in the mentioned paper. Also, we consider a two-parameter class of entropies derived from the maxent principle
proposed in [\textit{Kaniadakis, G. and Lissia, M. and Scarfone,
AM, Physica A: Statistical Mechanics and its Applications, 340(1)}]. We rederived this class by changing initial condition, obtaining the same class as our class derived from TGSK axioms.
\end{abstract}

\end{frontmatter}

\section{Two-parameter deformed entropy derived by TGSK axioms}
\label{wada_suyari_section}

Let $\Delta_n$ be an $n$-dimensional
simplex
\begin{equation}
   \Delta_n \equiv \left\{ (p_1, \dots , p_n) \Big\vert \; p_i \ge 0,
      \sum_{i=1}^{n} p_i = 1 \right \},
  \label{Delta}
\end{equation}
and let the set $R_{\alpha\beta}$ be given with
\begin{equation}
\label{ab-regions} R_{\alpha\beta}=R_\alpha \cup R_\beta,
\end{equation}
where
\begin{align*}
%&R_\alpha = [1, \infty] \times [0,1] \setminus (1,0),\nonumber\\
%&R_\beta =[0,1] \times [1, \infty] \setminus (0,1)
R_\alpha =  \Big\{ (\alpha, \beta) \in \sR^2 &\vert \;
           \alpha \ge 1,\ 0 \le \beta \le 1, (\alpha,\beta)\neq(1,0)   \Big\},\\
R_\beta =  \Big\{ (\alpha, \beta) \in \sR^2 &\vert \;
           0 \le \alpha \le 1,\ \beta \ge 1, (\alpha,\beta)\neq(0,1)
           \Big\}.
\end{align*}

Let $\sab$ denote a function $\sab: [0, 1] \to \set{R}$. For a generalized entropy, $\Sab: \Delta_n \to \set{R^+}$,
$(\alpha,\beta) \in R_{\alpha\beta}$, $n \in \sN$, of trace form
\begin{equation}
\Sab(p_1, p_2,...,p_n) = \sum_{i=1}^n \sab(p_i),
\label{traceForm}
\end{equation}
the following two-parameter generalized Shannon-Khinchin axioms are defined in \cite{wada2007two}:
\begin{itemize}
  \item {[TGSK1]} {\it continuity}:
         $\Sab$ is continuous in $\Delta_n$ and with respect to $\alpha$ and $\beta$;
  \item {[TGSK2]} {\it maximality}:
        for any $n \in N$ and any $(p_1, \dots, p_n) \in \Delta_n$
\begin{equation}
    \Sab(p_1, \dots, p_n) \le \Sab(\frac{1}{n},\dots ,\frac{1}{n})
\end{equation}
  \item {[TGSK3]} {\it two-parameter generalized Shannon additivity}:
if
\begin{align}
  p_{ij} \ge 0, \quad &p_i \equiv \sum_{j=1}^{m_i} p_{ij}, \quad
  p(j \vert i) \equiv \frac{p_{ij}}{p_i}, \nonumber \\
  &\forall i=1,\dots, n, \quad
  \forall j=1, \dots, m_i,
\end{align}
then the following equality holds:
\begin{align}
 \sum_{i=1}^{n} \sum_{j=1}^{m_i} \sab(p_{ij})
  &= \sum_{i=1}^{n} p_i^{\alpha} \sum_{j=1}^{m_i}
          \sab\left(p(j \vert i) \right)
\nonumber \\
   &+ \sum_{i=1}^{n} \sab(p_i)
         \sum_{j=1}^{m_i} p(j \vert i)^{\beta}.
  \label{TGSK3}
\end{align}
  \item {[TGSK4]} {\it expandability}:
      \begin{equation}
        \Sab(p_1, \dots, p_n, 0) = \Sab(p_1,\dots ,p_n).
      \end{equation}
\end{itemize}

\begin{theorem}
Let $S_{\alpha, \beta}: \Delta_n \ra \sR^+$, $(\alpha, \beta) \in R_{\alpha \beta}$, $n \in \sN$ be a function of trace form (\ref{traceForm}), which is not identically equal to zero for $n>1$ and which
satisfies [TGSK1]-[TGSK4]. Then, $S_{\alpha, \beta}: \Delta_n \ra
\sR^+$, $n \in \sN$, is uniquely determined with
\begin{equation}
   \Sab(p_1, \dots, p_n)
      = \sum_{i=1}^n \frac{p_i^{\alpha}-p_i^{\beta}}{\Cab},
   \label{Sab_entropy_definition}
\end{equation}
and $\Cab$ satisfies the following properties I)-IV):
\begin{itemize}
  \item[I)] $\Cab$ is continuous w.r.t. $\alpha$ and $\beta$,
     and has the same sign as $\beta-\alpha$.
     Consequently, $\Cab$ is antisymmetric under the interchange of $\alpha$
  and $\beta$, i.e., $\Cba = -\Cab$;
  \item[II)] $\lim_{\alpha \to \beta} \Cab  = 0$, and
         $\Cab \ne 0$ for $\alpha \ne \beta$;
  \item[III)] there exists an interval $(a, b) \in R$ such that $\Cab$ is
     differentiable w.r.t. both $\alpha$ and $\beta$ on the interval
     $(a, 1) \cup (1,b)$;
  \item[IV)] there exists a positive constant $k$ such that

 $\lim_{\alpha \to 1} \frac{d \Cab}{d \alpha} = -\frac{1}{k}$, and
            $\lim_{\beta \to 1} \frac{d \Cab}{d \beta} = \frac{1}{k}$.
\end{itemize}
\end{theorem}

\begin{remark} For $\alpha=1$ and $\beta=1,$ [TGSK1]-[TGSK4] reduce to original Shannon-Khinchin axioms, which uniquely determined Shannon entropy
\cite{khinchin1957mathematical}:
\begin{equation}
S_{1,1}\mathop = - k\sum\limits_{i=1}^n {p_i\ln p_i},
\label{Shannonentropy}
\end{equation}
where $k>0.$ Because of [TGSK1], we have
\begin{equation}
\ablim S_{\alpha, \beta} = S_{1,1} \mathop = - k\sum\limits_{i=1}^n {p_i\ln p_i}.
\label{Shannonentropy_lim}
\end{equation}
\end{remark}

According to proof from \cite{wada2007two}, form
(\ref{Sab_entropy_definition}) is uniquely determined by solving
functional equation (\ref{TGSK3}) from [TGSK3] and by using the
continuity with respect to $\Delta_n$ as assumed in [TGSK1]. Property (II) is sufficient and necessary for continuity with
respect to $\alpha$ and $\beta$ when $\alpha \neq \beta$.
Properties (III)-(IV) were required to ensure continuity in
$(\alpha, \beta) = (1, 1)$, so that in the limit case $S_{\alpha,
\beta}$ reduces to standard BGS entropy, i.e. to satisfy the property
defined by equation (\ref{Shannonentropy_lim}). The role of those
properties was to enable application of l'Hopital's rule to
(\ref{Sab_entropy_definition}).

Here, we follow a different approach and use the equality
\begin{align}
\label{theo: aux limit}
\lim_{\substack{x \to 0 \\ y \to 0}} \frac{p^x - p^y}{x - y}&=
\lim_{y \to 0} p^y \cdot%
\lim_{\substack{x \to 0 \\ y \to 0}} \frac{p^{x-y} - 1}{x -
y}\nonumber\\
&= \lim_{\substack{t \to 0}} \frac{p^t - 1}{t}= \ln p.
\end{align}
Accordingly, we have:
\begin{align}
&\ablim \Sab(p_1,\dots ,p_n) = \ablim \sum_{i=1}^n \frac{p_i^{\alpha} - p_i^{\beta}}{\Cab} \nonumber  \\
&= \ablim \sum_{i=1}^n \frac{\alpha - \beta}{\Cab} \cdot \frac{p_i^{\alpha} - p_i^{\beta}}{\alpha - \beta} \nonumber  \\
%&= \ablim \frac{\alpha - \beta}{\Cab} \cdot \sum_{i=1}^n p_i \cdot \frac{p_i^{\alpha-1} - p_i^{\beta-1}}{(\alpha-1) - (\beta-1)} \nonumber \\
&= \ablim \frac{\alpha - \beta}{\Cab} \cdot \ablim \sum_{i=1}^n p_i \cdot \frac{p_i^{\alpha-1} - p_i^{\beta-1}}{(\alpha-1) - (\beta-1)} \nonumber \\
&= -k \cdot \sum_{i=1}^{n}p_i\ln p_i
\label{limit_derivation}
\end{align}
and, therefore, properties (III) and (IV) should be replaced by
\begin{itemize}
  \item[III$^\prime$)] $\frac{C_{\alpha,\beta}}{\alpha - \beta}$ is continuous in $(1,1)$ and
\begin{equation}
\ablim \frac{\Cab}{\alpha - \beta} = -\frac{1}{k}.
\label{C_condition}
\end{equation}
\end{itemize}
Equivalently, the function $\Cab$ should be differentiable only in
$(\alpha,\beta)=(1,1)$, but need not be differentiable in a
neighbourhood of $(\alpha,\beta)=(1,1)$, as required by properties
(III) and (IV).

%\subsection{Counterexamples}

\section{Counterexamples to Wada-Suyari's theorem}

In this section we show that:

\begin{enumerate}[a)]
\item there exists a function $\Sab$ which does not belong to Wada-Suyari class
but has the form (\ref{Sab_entropy_definition}), with properties
(I),(II) and (III$^\prime$) satisfied, which means
that the conditions given by Wada and Suyari are not necessary for
satisfaction of the axioms [TGSK1]-[TGSK4], and
\item there exists a function $\Sab$ which belongs to Wada-Suyari class
but the limit $\lim_{\alpha \rightarrow 1,\beta \ra 1} \Sab$ does
not exist, which means that the conditions given by Wada and
Suyari are not sufficient for satisfaction of [TGSK1]-[TGSK4].
\end{enumerate}

\textbf{Counterexample a:}

The Weierstrass function is a well known example of nowhere
differentiable continuous function \cite{hardy1916weierstrass}. It
is defined with:
\begin{equation}
W(x)=\sum_{k=0}^\infty a^k \cos\left(b^k \pi x \right),
\end{equation}
where $0<a<1$ , $b$ is a positive odd integer, $ab > 1 + 3\pi / 2$
and $x \in \sR$. The Weierstrass function is bounded, since
\begin{equation}
\label{counterexample: weierstrass bound} |W(x)| \leq
\sum_{k=0}^\infty a^k | \cos\left(b^k \pi x \right)| \leq
\sum_{k=0}^\infty a^k = W(0) < \infty,
%= {1 \over 1 - a}
\end{equation}
where $W(0)=1 / (1-a)$. %Accordingly, $W(x) + 2 / (1-a) \neq 0$
Using the Weierstrass function we construct $\Cab(\alpha,\beta)$,
which satisfies properties (I), (II) and (III$^\prime$), but not
properties (III) and (IV).

Let
\begin{equation}
\label{counterexample: phi}
\Cab(\alpha,\beta) = \frac{1 - \alpha}{k} \cdot \frac{W(\alpha - 1) + 2 \cdot W(0)}{3 \cdot W(0)}.
\end{equation}
Since $W(x)$ is continuous and $W(x) + 2W(0) > 0$ according to
(\ref{counterexample: weierstrass bound}), $\Cab(\alpha,\beta)$
satisfies properties (I) and (II). Moreover,
\begin{equation}
\lim_{\substack{\alpha \to 1 \\ \beta \to 1}} \frac{\Cab}{\alpha -
\beta} = %
\lim_{\alpha \to 1} \frac{\Cab}{\alpha - 1} = -\frac{1}{k}
\label{C_condition_CounterB}
\end{equation}
and function $\Cab(\alpha,\beta)$ satisfies property
(III$^\prime$),

However, function $\Cab(\alpha,\beta)$ does not satisfy property
(III) from Wada-Suyari theorem since it is differentiable with
respect to $\alpha$ only in $\alpha=1$. Oppositely, the function
\begin{equation}
\frac{1}{\alpha-1} \cdot \Cab(\alpha,\beta) = \frac{1}{k} \cdot \frac{W(\alpha - 1) + 2 \cdot W(0)}{3 \cdot W(0)}
\end{equation}
should be differentiable for some $\alpha \neq 1$ as a product of
differentiable functions, further implying differentiability of
$W(\alpha - 1)$, which is impossible since the Weierstrass
function is nowhere differentiable.

\textbf{Counterexample b}

Let $C_{\alpha\beta}:R_{\alpha\beta}
\rightarrow \sR$
%\begin{equation}
%\Cab =
%\begin{cases}
%\frac{\alpha-\beta}{2k}  \cdot \left( %
%\frac{(\alpha - 1)(\beta - 1)}{(\alpha - 1)^2 + (\beta - 1)^2} -
%1\right) \quad & (\alpha, \beta) \in R_{\alpha\beta} \setminus
%(1,1)\\
%0 \quad &(\alpha,\beta) = (1,1)
%\end{cases}
%\end{equation}
\begin{equation}
\label{countex: Cab1} \Cab =
\frac{\alpha-\beta}{2k}  \cdot \left( %
\frac{(\alpha - 1)(\beta - 1)}{(\alpha - 1)^2 + (\beta - 1)^2} -
1\right)
\end{equation}
if $(\alpha, \beta) \neq (1,1)$ and
\begin{equation}
C_{1,1}=0
\end{equation}
Note that for $(x,y) \in \sR^2 \setminus (0,0)$
\begin{equation}
\label{countex: inequality}  -\frac{1}{2} \leq \frac{xy}{x^2 +
y^2} \leq \frac{1}{2},
\end{equation}
which follows from $(x \pm y)^2 \geq 0$. If we set $x=\alpha-1$
and $y=\beta-1$ in (\ref{countex: inequality}), it follows that
the term in brackets in expression (\ref{countex: Cab1}) is always
negative, which further implies that $\Cab$ has the same sign as
$\beta-\alpha$. In addition, for $(\alpha, \beta) \in
R_{\alpha\beta} \setminus (1,1)$, the function $\Cab$ is
continuous as composition of continuous functions,
while for $(\alpha,\beta)=(1,1)$ it is continuous since $\alpha -
\beta \rightarrow 0$ for $\alpha,\beta \rightarrow 0$ and the term
in brackets is bounded. Accordingly, $\Cab$ satisfies property
I. In addition, properties II-IV are straightforwardly
satisfied.

On the other hand, $\lim_{\alpha \rightarrow 1,\beta \rightarrow
1} \Cab/(\alpha - \beta)$ does not exist, since the expression
\begin{equation}
\label{countex: ab fract}
%\ablim \frac{\Cab}{\alpha - \beta}=%
%\frac{1}{2k}  \left(
%\ablim  %
\frac{(\alpha - 1)(\beta - 1)}{(\alpha - 1)^2 + (\beta - 1)^2} %-
%1\right)
\end{equation}
has no unique limit when $(\alpha,\beta)$ approaches (1,1) over different directions, which follows directly from
\begin{align}
\lim_{\substack{x \to 0 \\ y = k x}} \frac{xy}{x^2+y^2} = \lim_{x \to 0} \frac{x \cdot kx}{x^2+(kx)^2} = \frac{k}{1+k^2}.
\end{align}

%For example, if we take direction $\alpha = 1$, $\beta = t$, $+\infty > t > 1$, the expression (\ref{countex:ab fract}) tends to 0. On the other hand  if we take direction $\alpha = t + 1$, $\beta = 1 - t$, $ 1 > t > 0$, the expression (\ref{countex: ab fract}) tends to $-1/2$.
According to (\ref{limit_derivation}), the entropy $\Sab$ also has no limit when $(\alpha, \beta) \rightarrow (1,1)$, despite its belonging to Wada-Suyari class. On the other hand, the continuity of $\Cab/(\alpha - \beta)$ in $(1,1)$ is explicitly required by our property (III$^\prime$), which means that the corresponding entropy does not belong to our class.

%\section{Two parameter deformation by maximum entropy principle}
\section{Two-parameter deformed entropy derived by maximum entropy principle}
\label{kaniadakis_section}

Kaniadakis et. al. \cite{kaniadakis2004deformed} considered the
following class of trace-form entropies (in this work $k_{\rm
B}=1$)
\begin{equation}
S(p_1, \dots, p_n)=-\sum_{i=1}^n p_{_i}\,\Lambda(p_{_i}) \
,\label{defentropy}
\end{equation}
where $(p_1, \dots, p_n) \in \Delta_n$ represents a discrete
probability distribution, and $\Lambda(x)$ is an analytical
function that generalizes the logarithm. The canonical
distribution $(p_1, \dots, p_n)$ is obtained by maximizing the
entropy in equation (\ref{defentropy}) for fixed normalization and
energy, by requiring that the solution is represented with
generalized exponential ${\mathcal E}(x)=\Lambda^{-1}(x)$.
%
%Inspired by the maximum entropy principle, one of the bases of
%classical statistical mechanics, the authors in
%\cite{kaniadakis2004deformed}
%
%considered the following class of trace form entropies:
%\begin{equation}
%S(p)=-k \sum_{i=1}^N p_{_i}\,\Lambda(p_{_i}) , \label{defentropy}
%\end{equation}
%where $p\equiv\{p_{_i}\}_{_{i=1,\cdots,\,N}}$ is a discrete
%probability distribution, and $\Lambda(x)$ is an analytical
%function that generalizes the logarithm. By applying the maximum entropy
%principle to the previous class of entropies,
The following functional form for logarithm is derived:
\begin{equation}
\Lambda(x)=A_{_1}(\kappa_{_1},\,\kappa_{_2})\,
x^{\kappa_{_1}}+A_{_2}(\kappa_{_1},\,\kappa_{_2})\,
x^{\kappa_{_2}}\, \label{sol1},
\end{equation}
where $A_{_i}(\kappa_{_1},\,\kappa_{_2})$ are integration
constants which should be determined from the initial conditions
and from the continuity property of entropy. The following initial
conditions:
\begin{align}
\Lambda(1)&=0, \label{norm_first}   \\
\frac{d\,\Lambda(x)}{d\,x}\Bigg|_{x=1}&=1
\label{norm_second}
\end{align}
are used for all $\kappa_{_1},\,\kappa_{_2}$. From condition
(\ref{norm_first}), it follows that
\begin{equation}
A_1(\kappa_1, \kappa_2) = -A_2(\kappa_1, \kappa_2) =
A(\kappa_1, \kappa_2), \label{parameter_B}
\end{equation}
and the logarithm takes the form
\begin{equation*}
\Lambda(x) = A(\kappa_1,
\kappa_2) \cdot (x^{\kappa_1} - x^{\kappa_2});
\end{equation*}
and if the condition (\ref{norm_second}) is used, than we have
\begin{equation}
\Lambda(x)=\frac{x^{\kappa_{_1}}-x^{\kappa_{_2}}}{\kappa_{_1}
-\kappa_{_2}} \ .\label{log1}
\end{equation}
Conditions (\ref{norm_first}) and (\ref{norm_second}) are imposed
with the intention that $\Lambda(x)$ reduces to $\ln(x)$ and
$S(p)$ reduces to standard Shannon entropy in the limit case when
$\kappa_1$ and $\kappa_2$ approach 0. However, to accomplish
the same goal, it seems more natural to keep condition
(\ref{norm_first}) and require
\begin{equation}
\lim_{\substack{\kappa_1 \to 0 \\ \kappa_2 \to 0}} \Lambda(x) =
\ln(x) \label{norm_second_primed}
\end{equation}
instead of condition (\ref{norm_second}).
%From equations
%(\ref{sol1}) and (\ref{parameter_B}), it follows
%\begin{equation*}
%\Lambda(x) = \frac{x^{\kappa_1} - x^{\kappa_2}}{B(\kappa_1,
%\kappa_2)}.
%\end{equation*}
By using limit condition (\ref{norm_second_primed}), we obtain
\begin{align*}
\ln x &= \lim_{\substack{\kappa_1 \to 0 \\ \kappa_2 \to 0}}
\Lambda(x)
       =  \lim_{\substack{\kappa_1 \to 0 \\ \kappa_2 \to 0}} A(\kappa_1, \kappa_2) \cdot (x^{\kappa_1} - x^{\kappa_2})      \\
       &= \lim_{\substack{\kappa_1 \to 0 \\ \kappa_2 \to 0}}  A(\kappa_1, \kappa_2) \cdot ({\kappa_1} - {\kappa_2}) \cdot \frac{x^{\kappa_1} - x^{\kappa_2}}{\kappa_1 - \kappa_2}     \\
       &= \lim_{\substack{\kappa_1 \to 0 \\ \kappa_2 \to 0}}  A(\kappa_1, \kappa_2) \cdot ({\kappa_1} - {\kappa_2}) \cdot
       \lim_{\substack{\kappa_1 \to 0 \\ \kappa_2 \to 0}}  \frac{x^{\kappa_1} - x^{\kappa_2}}{\kappa_1 - \kappa_2}     \\
%       &= \lim_{\substack{\kappa_1 \to 0 \\ \kappa_2 \to 0}} \frac{x^{\kappa_1} - x^{\kappa_2}}{\kappa_1 - \kappa_2} \cdot \lim_{\substack{\kappa_1 \to 0 \\ \kappa_2 \to 0}} \frac{\kappa_1 - \kappa_2}{C(\kappa_1, \kappa_2)}     \\
%       &= \ln x \cdot \frac{1}{k} \cdot \lim_{\substack{\kappa_1 \to 0 \\ \kappa_2 \to 0}} \frac{\kappa_1 - \kappa_2}{C(\kappa_1, \kappa_2)},    \\
\end{align*}
and by using equality (\ref{theo: aux limit}) limit condition (\ref{norm_second_primed}) becomes
condition
\begin{equation}
\label{maxent: A_limit_condition}
\lim_{\substack{\kappa_1 \to 0 \\ \kappa_2 \to 0}}  A(\kappa_1, \kappa_2) \cdot ({\kappa_1} - {\kappa_2}) = 1.
\end{equation}
Note that limit condition (\ref{norm_second_primed}) is equivalent to the following modification of condition (\ref{norm_second}):
\begin{equation}
\lim_{\substack{\kappa_1 \to 0 \\ \kappa_2 \to 0}} \frac{d\,\Lambda(x)}{d\,x}\Bigg|_{x=1}=1.
\end{equation}

Now, by imposing condition (\ref{maxent: A_limit_condition}), entropy (\ref{defentropy}) takes the form
\begin{equation}
   S(p) = - k \cdot \sum_{i=1}^n A(\kappa_1, \kappa_2)\cdot (p_i^{\kappa_1+1}-p_i^{\kappa_2+1}).
   \label{Sab}
\end{equation}
If we introduce $\kappa_1 = \alpha - 1$, $\kappa_2 = \beta - 1$ and
\begin{equation}
%\label{maxent: Cab}
C(\alpha, \beta) = - \frac{1}{k \cdot
A(\alpha-1, \beta - 1)},
\end{equation}
the functional form of the entropy  is the same as the one derived
by the functional equation given by [TGSK3]. In addition, if the
continuity condition given by [TGSK1] is imposed, condition
(\ref{maxent: A_limit_condition}) reduces to the conditions [I],
[II] and [III$^\prime$]. Accordingly, the entropy class of
entropies (\ref{defentropy}) is the same as the one proposed in
section \ref{wada_suyari_section}.

%\section{References}

\bibliographystyle{IEEEtran}
%\bibliography{IEEEabrv}
\bibliography{reference}

\end{document}